\documentclass[twocolumn,showpacs,amsmath,amssymb,prl,nobibnotes,floatfix
]{revtex4-1}

\usepackage{bm}
\usepackage[plainpages=false]{hyperref}
\usepackage{graphicx}
\usepackage{multirow}
\usepackage{rotating}
\usepackage[utf8]{inputenc}

\usepackage{color}
\usepackage[normalem]{ulem}

\usepackage{dcolumn}
\newcolumntype{d}[1]{D{.}{.}{#1}}

\def\fun#1#2{\lower3.6pt\vbox{\baselineskip0pt\lineskip.9pt
  \ialign{$\mathsurround=0pt#1\hfil##\hfil$\crcr#2\crcr\sim\crcr}}}

\newcommand{\cmmnt}[1]{}

\newcommand{\mnras}{Mon.\ Not.\ R.\ Astron.\ Soc.\ }
\newcommand{\aap}{Astron.\ Astrophys.\ }
\newcommand{\jcap}{J.\ Cosmol.\ Astropart.\ Phys.\ }
\newcommand{\physrep}{Phys.\ Rep.\ }

\newcommand{\apjs}{Astrophys.\ J.\ S.\ }
\newcommand{\aaps}{Astron.\ Astrophys.\ Suppl.\ Ser.\ }

\begin{document}

\title{Axion Condensate Dark Matter Constraints from Resonant Enhancement of Background Radiation}

\author{G\"unter Sigl$^{1}$}
\email[]{guenter.sigl@desy.de}

\author{Pranjal Trivedi$^{1,2,3}$}
\email{pranjal.trivedi@desy.de}

\affiliation{$^{1}$Universit\"at Hamburg, {II}. Institut f\"ur Theoretische Physik, Luruper Chaussee 149, 22761 Hamburg, Germany.\\
$^{2}$Hamburger Sternwarte, Gojenbergsweg 112, 21029 Hamburg, Germany.\\
$^{3}$Department of Physics, Sri Venkateswara College, University of Delhi 110020 India}

\begin{abstract}
We investigate the possible parametric growth of photon amplitudes in a background of axion-like particle (ALP) dark matter.
The observed extragalactic background radiation limits the allowed enhancement effect. We derive the resulting constraints on the axion-photon coupling constant $g_{a\gamma}$ from Galactic ALP condensates as well as over-densities.
If ALP condensates of size $R$ exist in our Galaxy, a scan for extremely narrow unresolved spectral lines with frequency $\nu$ can constrain the axion-photon coupling at ALP mass $m_a=4\pi\nu$ to $g_{a\gamma}\lesssim2\times10^{-14}(10\,{\rm kpc}/R)\,{\rm GeV}^{-1}$. 
Radio to optical background data yield constraints at this level within observed wavebands or ALP mass windows over a broad range $0.08\,  \mu \text{eV} \lesssim m_a \lesssim 8 \text{ eV}$. 
These condensate constraints on $g_{a\gamma}$ probe down to the QCD axion band for $m_a \gtrsim 10 \mu$  eV.
\end{abstract}

\maketitle

\textit{Introduction.--}
One of the leading candidates for dark matter \cite{Bertone:2016nfn,Salucci:2018hqu,Lin:2019uvt} are axion-like particles (ALPs) which correspond to pseudoscalar fields $a$. They possess a two-photon coupling of the form $g_{a\gamma}\,a\,F_{\mu\nu}\tilde F^{\mu\nu}/4$, which is the most relevant coupling in dilute media. 
ALPs are generalizations of axions, originally motivated to solve the strong charge-parity (CP) problem by means of promoting the CP-violating phase $\theta$ to $a/f_a$ where $f_a$ is known as the Peccei-Quinn scale \cite{Peccei:1977PhRvL..38.1440P,Weinberg:1978PhRvL..40..223W,Wilczek:1978PhRvL..40..279W}. 
Through its couplings to gluons and quarks the axion attains a mass $m_a$ below the quantum chromodynamics (QCD) scale and its expectation value is driven to zero. ALPs also arise generically in low energy effective field theories of string compactifications. \cite{Svrcek:2006yi,Conlon:2006tq,Arvanitaki10:PhysRevD.81.123530,Cicoli:2012sz}.

In addition to their coupling to photons $g_{a\gamma}$ ,
ALPs are characterized through
their vacuum mass $m_a$. ALPs generally do not solve the strong CP problem, unlike axions, and $g_{a\gamma}$ and $m_a$ are taken as  as independent parameters. 
The axion-photon coupling term leads to ALP-photon oscillations in the presence of external electromagnetic fields. 
It also leads to an effective refractive index for photons propagating in a background of ALPs as well as parametric growth of an impinging photon beam.
While the former effect has been investigating extensively in both cosmological, astrophysical
contexts (for reviews see Ref.~\cite{JaeckelRingwald2010:2010ni,Arias:2012az,Marsh:2015xka}) and in experimental approaches (for a review see Ref.~\cite{Graham:2015ouw,Irastorza:2018dyq}), the latter so far is still less well studied. 
Parametric growth of photon amplitudes and refractive effects can be particularly relevant if ALPs constitute a significant part of the dark matter \cite{Preskill:1982cy,Abbott:1982af,Dine:1982ah,VisinelliGondolo2009:PhysRevD.80.035024,Marsh:2015xka}which is what we assume here without specifying the ALP production processes. 
In Ref.~\cite{Sigl:2018fba} we have studied the birefringent effect of ALP dark matter on the cosmic microwave background (CMB) which leads to strong constraints on $g_{a\gamma}$ in the mass range
$10^{-27}\,{\rm eV}\lesssim m_a\lesssim10^{-22}\,{\rm eV}$ which overlaps with the mass range of fuzzy dark matter \cite{Hu2000FuzzyCDM:PhysRevLett.85.1158,Hui:2017PhRvD..95d3541H}.

In this Letter, we investigate the possible parametric growth of diffuse background photons impinging on ALP dark matter condensate as well as ALP over-densities. 
We find that avoiding the overproduction of background radiation leads to strong constraints on $g_{a\gamma}$ for condensates over a wide range of ALP masses. 
Constraints in the mass range of micro electron volts are also obtained on the mass and size distributions of ALP dark matter over-densities. 
The parametric enhancement of photons we describe is independent of Galactic or cosmic magnetic fields and distinct from ALP-photon conversion \cite{Raffelt:1987im,Kelley:2017vaa,Sigl:2017sew,Mukherjee:2018oeb}.

The parts of the Lagrangian depending on the ALP and photon fields can be written as
\begin{equation}\label{eq:L_a}
  {\cal L}_{a\gamma}=-\frac{1}{4}F_{\mu\nu}F^{\mu\nu}+\frac{1}{2}\partial_\mu a\partial^\mu a+
  \frac{1}{4}g_{a\gamma}\,a\,F_{\mu\nu}\tilde F^{\mu\nu}-V_a(a)\,,
\end{equation}
using Lorentz-Heaviside units $\epsilon_0=\mu_0=1$ and natural units $c=\hbar=k_B=1$.
Here, $F_{\mu\nu}$ is the electromagnetic field strength tensor, $\tilde F_{\mu\nu}$ is its dual and
$C_{a\gamma}$ is a model dependent dimensionless parameter. The
effective ALP potential $V_a(a)$ can be expanded as $V_a(a)=\frac{1}{2}m_a^2a^2+{\cal O}(a^3)$ around $a=0$,
with $m_a$ the effective ALP mass. The axion-photon coupling constant can be written as
\begin{equation}\label{eq:f_a}
  g_{a\gamma}=\frac{s\alpha_{\rm em}}{2\pi f_a}\,,
\end{equation}
where $s$ is a model-dependent parameter of order unity, $\alpha_{\rm em}$ the fine structure constant.

\textit{Photon Propagation in an ALP background.--} \label{sec:photon_propagation}
Considering left- and right-circular polarization photon modes propagating in the $z-$direction, we make the Ansatz
\begin{equation}\label{eq:A+-}
  {\bf A}_\pm(t,{\bf r})=A_\pm(t){\bf e}_\pm e^{ikz}\,,
\end{equation}
where ${\bf e}_\pm\equiv{\bf e}_x\pm i{\bf e}_y$ are the left and right-circular mode unit vectors.
To zeroth order, photon wave-packets will propagate along trajectories $z=t$ enabling us to
identify time and length scales from here on. For a monochromatic ALP field, Eq.~(\ref{eq:A+-}) then yields the equation of motion
\begin{equation}\label{eq:Mathieu}
  \left(\frac{\partial^2}{\partial t^2}+k^2\right)A_\pm(t)=\pm\frac{km_ag_{a\gamma}}{\epsilon_0}a_0\cos(m_at+\delta)A_\pm(t)\,.
\end{equation}
Here, $a_0$ is the amplitude of the ALP field which is supposed to vary on time and lengths scales much larger
than $1/k$ and the inverse photon frequency. The random phase $\delta$ changes on the length scale of the coherence length $l_c$ of the ALP field. Eq.~(\ref{eq:Mathieu}) has the form of a Mathieu equation which can be brought into standard form (up to the phase $\delta$)
\begin{equation}\label{eq:Mathieu2}
  \left[\frac{d^2}{d x^2}+A-2q\cos(2x+\delta)\right]y(x)=0
\end{equation}
via the substitutions
\begin{eqnarray}
  x&\equiv&m_at/2\,,\,
  A=\frac{4k^2}{m_a^2}\,,\,\,\,\label{eq:Mathieu2a}
  q=\pm\frac{2kg_{a\gamma}}{\epsilon_0}\frac{a_0}{m_a}\simeq\pm\,4.3\times10^{-20}\nonumber \\
  &\times& \!\!\left(g_{a\gamma}10^{14}\,{\rm GeV}\right)
  \!\left(\frac{\mu{\rm eV}}{m_a}\right)
  \!\left(\frac{\rho_a}{0.3\,{\rm GeV}{\rm cm}^{-3}}\right)^{\!1/2}
  \!\left(\frac{k}{m_a}\right)\!\!.
\end{eqnarray}
For the ALP amplitude $a_0$ we utilize the relation
\begin{equation}\label{eq:a0}
  \rho_a=\frac{1}{2}m_a^2a_0^2\,,
\end{equation}
with $\rho_a$ the local ALP dark matter energy density. In the supplementary material we derive the properties of solutions to the Mathieu equation~(\ref{eq:Mathieu}) in the limit of $|q|\ll1$, relevant for photon propagation in an ALP background.

\textit{Constraint from Galactic ALP Condensates.--}
One can now obtain our most stringent constraint in the following way: If one believes that the observed radio fluxes
are mostly due to astrophysical processes, then the parametric resonance caused by the smooth dark matter component
should not significantly increase observed fluxes. As shown in the supplementary material, a parametric resonance occurs in the Mathieu
equation for $|1-A|\lesssim|q|/2$ which from Eq.~(\ref{eq:Mathieu2a}) corresponds to wavenumbers $|k-m_a/2|\lesssim m_a|q|/8$.
Thus the relative width of the resonance is $|q|/2$ and the intensity growth rate is $\simeq|q|m_a/\sqrt2$ and one has
\begin{equation}\label{eq:constr1a}
  \int_0^1\frac{d\tilde q}{2}\exp\left(m_a\int dl|q(l)|\Theta[|q(l)|-\tilde q]/\sqrt2\right)\lesssim f\frac{\Delta\nu}{\nu}\,,
\end{equation}
where $|q(l)|$ is the value of $|q|$ given by Eq.~(\ref{eq:Mathieu2a}) along the line of sight parametrized by the length $l$, $\Theta(x)$
is the Heaviside function which limits the line of sight integral to regions in which $|q|>\tilde q$, and $f>1\sim{\cal O}(1)$
is the possible enhancement consistent with the data. The factor $\Delta\nu/\nu$ appears if one scans with a frequency bandwidth
$\Delta\nu$ so that the received un-enhanced flux decreases proportional to $\Delta\nu$.
For example, if $\rho_a(r)$ is monotonously decreasing with the distance $r$
from the Galactic center, then according to Eq.~(\ref{eq:Mathieu2a}) there is an $r_c(\tilde q)$ such that $|q(r)|>\tilde q$ for $r<r_c(\tilde q)$
and the exponent in Eq.~(\ref{eq:constr1a}) will have the form $\int_0^{r_c(\tilde q)}dr|q(r)|$ and can be computed explicitly for a
given profile $\rho_a(r)$. For a rough estimate assuming that $q$ is constant along the line of sight of total length $R$ gives
\begin{equation}\label{eq:constr1b}
  |q|\exp\left(m_a|q|R/\sqrt2\right)\lesssim2f\frac{\Delta\nu}{\nu}\,.
\end{equation}
With Eq.~(\ref{eq:Mathieu2a}) this yields
\begin{equation}\label{eq:constr1c}
  g_{a\gamma}\lesssim1.9\times10^{-14}\!\left(\frac{10\,{\rm kpc}}{R}\right)
  \!\left(\frac{\rho_a}{0.3\,{\rm GeV}{\rm cm}^{-3}}\right)^{\!\!-1/2}{\rm \!GeV}^{-1}\,,
\end{equation}
using $k \simeq m_a /2, f \sim 1$ and $\Delta\nu/\nu \sim 1$.  This estimate neglects the additional weak logarithmic dependencies on deviations of $g_{a\gamma}$, $m_a$, $\rho_a$, $f$ and $\Delta\nu/\nu$
from their fudge values used above. 
Note that this constraint on $g_{a\gamma}$ only depends on the ALP density and Galactic scale $R$, but not on the ALP mass $m_a$ or the parameter $s$. 

\begin{widetext}

\begin{figure}
\centering
\includegraphics[width=0.83\columnwidth]{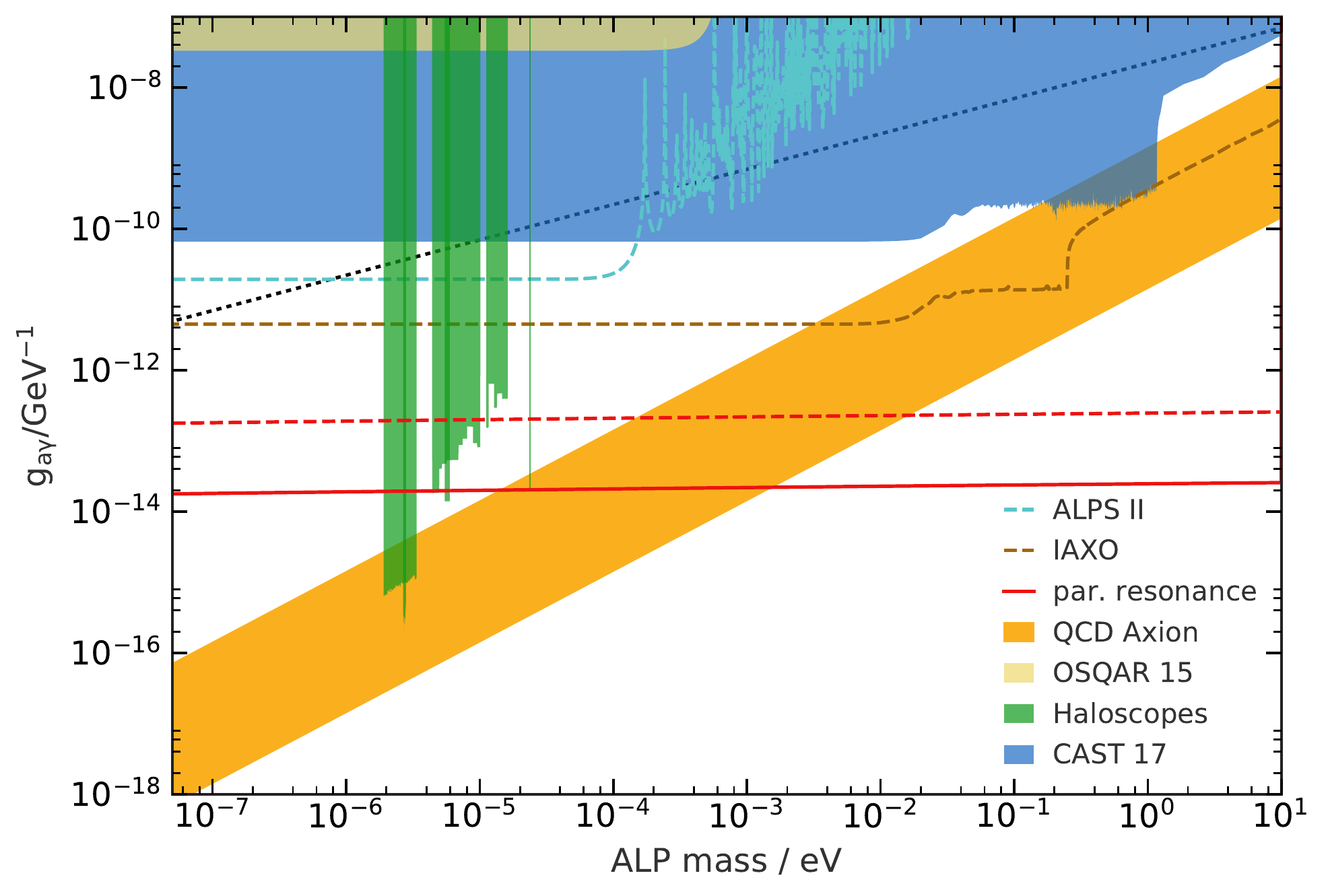}
\caption{
Axion-photon coupling constant $g_{a\gamma}$ as a function of ALP mass $m_a$. The red curves denote the constraint derived from parametric resonance due to axion condensate Galactic dark matter derived from Eq.~(\ref{eq:constr1b}), for $R=10$ kpc (solid) and for $R=1$ kpc (dashed). Resonant enhancement $f\sim1$ and relative channel bandwidth $\Delta \nu / \nu \sim 1$ are assumed. The constraint on $g_{a\gamma}$ scales logarithmically with $m_a$ (and logarithmically with $f \Delta \nu / \nu$, if they vary).
For comparison, excluded (filled) regions and forecasts (dashed lines) are shown from helioscope,  haloscope and light-shining-through-walls experiments. The allowed parameter space for temperature-dependent ALP cold dark matter via the misalignment mechanism \cite{Ringwald:2013via} is the region below the dotted black line. The orange parallel band depicts QCD axion models \cite{Kim:1979if,Shifman:1979if,Dine:1981rt,Zhitnitsky:1980tq}.
}
\label{fig:g_m_plot}
\end{figure}

\end{widetext}

A more precise numerical solution of  Eq.~(\ref{eq:constr1b}), shown in Fig.~\ref{fig:g_m_plot}, displays the slight logarithmic weakening of the $g_{a\gamma}$ constraint as a function of $m_a$, by $\sim$ 35$\%$ over the mass range. 
The lower end of the mass range appropriate for ALP parametric resonance is set by the lowest available radio frequency, $\nu \simeq$ 10 MHz or $ m_a \simeq 0.083 \mu$ eV. The upper end $m_a\simeq 10$ eV is determined by the condition that the non-relativistic ALP temperature remains below the critical condensate temperature,
\begin{equation}\label{eq:crit_cond_temp}
  T_c = \frac{2 \pi}{\zeta(3/2)^{2/3}} \frac{\rho_a^{2/3}}{m_a^{5/3}} \gtrsim T_{\text{virial}} \simeq \frac{1}{2} m_a v_a^2.
\end{equation}
We note that the constraint Eq.~(\ref{eq:constr1c}) likely only holds if there exist ALP condensates of size $R$ since the parametric
resonance is extremely narrow,
of relative width of order $|q|$ given by Eq.~(\ref{eq:Mathieu2a}). Therefore, for the resonance not to be washed out the ALP field
has to be essentially mono-energetic which requires a condensate. In practice this means that the zero mode should contribute
a significant fraction to the ALP density and should be described by a classical field whose amplitude $a(t,{\bf r})$ varies on time scales
much larger than $1/m_a$. In fact, adiabaticity requires that the rate at which the amplitude $a(t,{\bf r})$ varies should be smaller than
the resonant enhancement rate 
\begin{eqnarray}\label{eq:Rc}
  R_c&\simeq&\frac{m_a|q|}{\sqrt2}\simeq\frac{\sqrt2kg_{a\gamma}a_0}{\epsilon_0}\simeq\frac{g_{a\gamma}\rho_a^{1/2}}{\epsilon_0} \simeq2.3\times10^{-11}\nonumber \\ 
  &\times&\left(g_{a\gamma}10^{14}\,{\rm GeV}\right)\left(\frac{\rho_a}{0.3\,{\rm GeV}{\rm cm}^{-3}}\right)^{1/2}{\rm s}^{-1}\,,
\end{eqnarray}
where we have used Eq.~(\ref{eq:Mathieu2a}) and $k\simeq m_a/2$. Note that this is independent of $m_a$.
The time scales on which scalar field amplitudes evolve are determined by the hydrodynamical equations which are similar to WIMP dark matter but also include some extra terms, 
in particular quantum pressure. Above the Jeans scale, time evolution is roughly governed by free fall with a time scale $\sim R/v\sim10^3R$ at length scales $R$. For the smooth dark
matter component on galactic scale this is much longer than the inverse of Eq.~(\ref{eq:Rc}). On lines of sight crossing
small scale structure evolving with rates larger than Eq.~(\ref{eq:Rc}), corresponding to structures on length scales
$R\lesssim0.4(v/10^{-3})(g_{a\gamma}10^{14}\,{\rm GeV})^{-1}(0.3\,{\rm GeV}{\rm cm}^{-3}/\rho_a)^{1/2}\,$pc,
the condition of adiabaticity is likely violated such that constraints may not be easily derived from observations in such directions.
A precise description of cosmic and galactic structure formation with ALP dark matter is challenging \cite[e.g.][]{Sikivie:1997ngCaustics,Marsh:2015xka,SakharovKhlopov:1994id,Enander:2017ogx,Vaquero:2018tib,Veltmaat:2018dfz}.
It should also be kept in mind that the extent to which ALP condensates form is controversial~\cite{Sikivie:2009qn,Davidson:2013aba,Davidson:2014hfa}. 
The quantum evolution of a self-gravitating axion field can provide a limit on the lifetime of the condensate \cite{Chakrabarty:2017fkd}, which could be further modified in the presence of inhomogeneities. However this lifetime is very long compared to condensate formation timescales, for the mass range of interest, and also compared to the lifetime required from adiabaticity (Eq.~\ref{eq:Rc}). 

If $\rho_a(r)\propto r^{-\alpha}$
close to the galactic center, for $\alpha>2$ the constraint may become stronger since $|q|r\propto\rho_a^{1/2}r$. To this end
we integrate over an NFW or Burkert dark matter density profile (details in supplementary material)
to find that the constraint on $g_{a\gamma}$ can  tighten by a factor $\sim 2-3$ or $\sim 5$ when we integrate from the center till 10 kpc or 100 kpc, respectively.
An angular anisotropy in the enhanced background signal is also expected due to our offset from the Galactic center. the level of anisotropy depends on the scale $R$ and profile (details in supplementary material). 
This could be exploited to further tighten constraints on $g_{a\gamma}$ requiring the predicted signal to be consistent with the highly isotropic unresolved radio background \cite{Singal:2017jlh,Holder:2012nm}.

Note that telescopes measuring the diffuse background
spectrum would smear out the signal over frequency resolution of the instrument and since we calculated the total ALP energy fraction converted
to radio photons, they would still detect a significant enhancement at frequencies around $\nu\simeq m_a/(4\pi)$ if the limit Eq.~(\ref{eq:constr1c}) is violated. 
\begin{widetext}

\begin{figure}
\centering
\includegraphics[width=0.83\columnwidth]{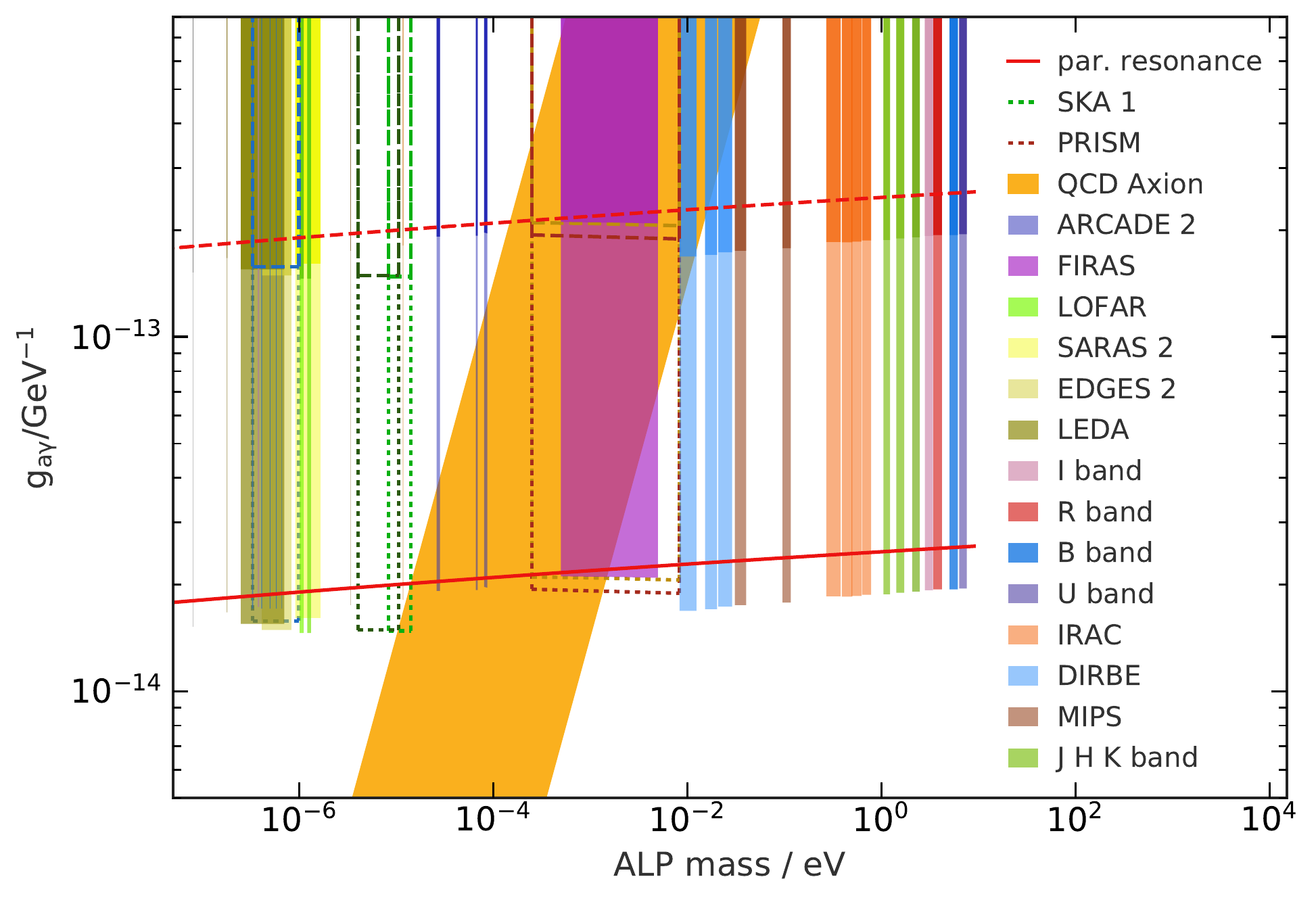}
\caption{Constraints on 
$g_{a\gamma}$ as a function of $m_a$ from parametric resonance due to axion condensate Galactic dark matter. Shown here is a narrower vertical sub-range $5\times10^{-15} < g_{a\gamma} < 8\times10^{-13}$ of Fig.~\ref{fig:g_m_plot} around our expected constraints (red curves, as defined for Fig.~\ref{fig:g_m_plot}).
Regions that can be excluded for condensates by current radio, infrared and optical background observations (filled) and future observations (lines) are shown. 
An enhancement factor $f = 1 + S_{\text{enh}}/S$ and a relative channel bandwidth $\Delta \nu / \nu$ were used to obtain each constraint on $g_{a\gamma}$ over a mass window given by the particular spectral waveband. 
Observational constraints also weaken by a factor of 10 from the case $R=10$ kpc (lighter shading and dotted lines) to the case $R=1$ kpc (darker shading and dashed lines).
}
\label{fig:g_m_zoom_plot}
\end{figure}

\end{widetext}

Furthermore, a scan in radio frequencies for such a line could strengthen the constraint for $m_a=4\pi\nu$, albeit only logarithmically in $\Delta\nu/\nu$.
By comparing to radio survey observations (and forecasts), detailed in the supplementary material, we can exclude several regions of parameter space, as shown and labelled in Fig.~\ref{fig:g_m_zoom_plot}.

Note that the sensitivity is not far from the QCD axion band
\begin{equation}\label{eq:QCD}
  g_{a\gamma}\sim2.4\times10^{-16}\left(\frac{m_a}{\mu{\rm eV}}\right){\rm GeV}^{-1}\,,
\end{equation}
for $m_a\gtrsim10\,\mu$eV (Fig.~\ref{fig:g_m_plot}). 
Note that QCD axion models, in post-inflationary PQ symmetry breaking, account for all cosmic dark matter over the `classic' axion window $7\times 10^{-5}$ eV $\lesssim m_a \lesssim 6 \times 10^{-3}$ eV \cite{Marsh:2015xka}, which our constraints are sensitive to (Fig.~\ref{fig:g_m_zoom_plot}).
Pre-inflationary PQ symmetry breaking, with minimal tuning of the axion initial displacement at the level of $10^{-1}$, implies an approximate QCD axion mass range of $ 10^{-6}$ eV $\lesssim m_a \lesssim 10^{-4}$ eV \cite{Irastorza:2018dyq}, which our constraints overlap at its higher mass end.  

\textit{Discussion}
In the supplementary material we consider, more generally, parametric enhancement in the presence of ALP over-densities. We
find that the most frequently discussed ALP structures, axion mini-clusters and axion stars, do not lead to significant constraints
compared to the effect of the average Galactic ALP density.
A similar study has been performed in Ref.~\cite{Arza:2018dcy} which obtains somewhat weaker sensitivities when assuming a non-coherent homogenous ALP field in a nearby caustic ring. 
Note that we integrate the exponential amplification factor rather than average it and our coupling constraint has a different mass dependence. 
In contrast to Ref.~\cite{Arza:2018dcy}, we do not get significant sensitivities from ALP stars and mini-clusters.
A conceptual study of ALP coupling to photon fields has been presented in Ref.~\cite{Sawyer:2018ehf} and photon emission from ALP over-densities has also been investigated in Ref.~\cite{Hertzberg:2018zte}.

We note that the constraints based on the spontaneous and induced ALP decay~\cite{Caputo:2018ljp,Caputo:2018vmy} are different 
because they correspond to the decay of single ALPs within a photon field rather than photon propagation in a high density ALP field.
To prevent overproduction of the radio background $\Omega_\gamma(\epsilon\sim m_a/2)$
requires that $f_{\rm dm}$ times the ALP fraction decaying during one Hubble time $t_H$ should be smaller than
the ratio of the radio photon to dark matter densities,
\begin{equation}\label{eq:constr2}
  f_{\rm dm}\frac{t_H}{\tau_a}(1+2f_\gamma)\lesssim\frac{\Omega_\gamma(m_a/2)}{\Omega_{\rm dm}}\,,
\end{equation}
where $\tau^{-1}_a$ is the spontaneous ALP decay rate which could be enhanced by a factor $(1+2f_\gamma)$
due to induced emission in an environment with average occupation number $f_\gamma$ at energy $\epsilon=m_a/2$
which could reach a few orders of magnitude~\cite{Caputo:2018ljp,Caputo:2018vmy}. Using $\tau^{-1}_a=g^2_{a\gamma}m_a^3/(64\pi)$
for radiative decays yields
\begin{eqnarray}\label{eq:constr2a}
  g_{a\gamma}&\lesssim& 5\times10^{-3}f_{\rm dm}^{-1/2}(1+2f_\gamma)^{-1/2} \nonumber \\ 
  &\times& \left(\frac{\Omega_\gamma(m_a/2)}{10^{-10}\Omega_{\rm dm}}\right)^{1/2} \left(\frac{\mu{\rm eV}}{m_a}\right)^{3/2}\,{\rm GeV}^{-1}\,.
\end{eqnarray}
While this constraint is not very strong at $\mu$eV masses, we note that for $m_a\sim\,$keV one has
$\Omega_\gamma(m_a/2)\sim10^{-7}\Omega_{\rm dm}$, so that the constraint reads $g_{a\gamma}\lesssim10^{-14}\,{\rm GeV}^{-1}$.

\textit{Conclusions} \label{sec:Conclusions}
We have investigated the possible parametric enhancement of the background photon flux propagating through ALP dark matter characterized by the ALP mass $m_a$ and its coupling to photons $g_{a\gamma}$. 
The equation relevant for parametric growth is a Mathieu-type equation and we have provided a general expansion of its solutions in the limit $|q|\ll1$, relevant for axion-photon coupling. 
We find dispersion quadratic in $q$ for $|A-1|\gtrsim|q|$ and parametric resonances with growth rates ${\cal O}(|q|)$ for $|A-1|\lesssim|q|$.
We find that exponential enhancement can occur along lines of sight which are dominated by a smooth
ALP component which is predominantly in a condensate state. 
The line of sight should not cross significant small-scale  ALP over-densities of size $R\lesssim0.4\,(v/10^{-3})(g_{a\gamma}10^{14}\,{\rm GeV})^{-1}(0.3\,{\rm GeV}{\rm cm}^{-3}/\rho_a)^{1/2}\,$pc.

Assuming that the observed background is mostly astrophysical, the enhancement should not be larger than factors of a few, and an in-principle constraint $g_{a\gamma} \lesssim 2 \times 10^{-14}$ GeV$^{-1}$ is found over $m_a \lesssim 10$ eV. Using existing radio, infrared and optical background observations (and forecasts), we can constrain several windows of the ALP mass range $0.08 \mu \text{ eV} \lesssim m_a \lesssim 8 \text{ eV}$ at a coupling $g_{a\gamma} \lesssim 1.5-2.1 \times 10^{-14}$ GeV$^{-1}$, for $R=10$ kpc. While based on different assumptions, ALP condensate limits on $g_{a\gamma}$ are two or more orders of magnitude stronger than those from helioscopes and light-shining-through-walls experiments and can cover a broader ALP mass range compared to haloscopes. For $m_a\gtrsim10\,\mu$eV the sensitivity can reach the QCD axion band.

In contrast, parametric conversion of ALP over-densities to photons is unlikely to significantly increase the diffuse photon background provided such ALP over-densities have characteristic
sizes and masses of order $R\sim1/m_a$ and $M\sim f_a^2/m_a$, respectively. 
Finally, we have shown that spontaneous and induced ALP decays into two photons can contribute significantly to diffuse photon fluxes only for masses $m_a$ of electronvolts and above.

\textit{Acknowledgments}: This work has been supported by the Deutsche Forschungsgemeinschaft through the Collaborative Research Center SFB 676 ``Particles,Strings and the Early Universe'' 
and under Germany's Excellence Strategy - EXC 2121 "Quantum Universe" - 39083306. 
We acknowledge useful conversations with Ariel Arza, Robi Banerjee, Volker Heesen, Shane O'Sullivan, Andreas Pargner, Georg Raffelt, Javier Redondo, Andreas Ringwald, Thomas Schwetz and Elisa Todarello. The figure preparation has made use of ALPlot \cite{Alplot}).





\appendix

\begin{widetext}

\section{SUPPLEMENTARY MATERIAL}

\section{The small $|q|$ expansion}
\label{sec:Mathieu}
On length and time scales $l$ in the range $1/m_a\lesssim l\lesssim l_c$ in Eq.~(\ref{eq:Mathieu})
one can make use of the Floquet theorem which states that solutions of Eq.~(\ref{eq:Mathieu2}) have the form
\begin{equation}\label{eq:Floquet}
  y(x)=e^{i\mu x}f(x)\,,
\end{equation}
where $f(x)$ is a function that is periodic with period $\pi$, i.e. $f(x+\pi)=f(x)$ and $\mu(A,q)$ is known as the Floquet exponent
which depends on $A$ and $q$.

To make this a bit more quantitative we now make the ansatz
\begin{equation}\label{eq:ansatz}
 y(x)=e^{i\alpha(x)}\,.
\end{equation}
Inserting into Eq.~(\ref{eq:Mathieu2}) yields the differential equation
\begin{equation}\label{eq:Mathieu_mod}
 i\alpha^{\prime\prime}(x)-\left(\alpha^\prime\right)^2+\left[A-2q\cos(2x+\delta)\right]=0\,,
\end{equation}
where a prime denotes a derivative with respect to $x$. We now write $\alpha^\prime$ as a Fourier series that
is periodic in $2x$,
\begin{equation}\label{eq:ansatz2}
 \alpha^\prime(x)=\sum_{n=-\infty}^{+\infty}c_ne^{2inx}\,.
\end{equation}
Note that the coefficient for $n=0$ is the Floquet exponent, $c_0=\mu$.
Substituting Eq.~(\ref{eq:ansatz2}) into Eq.~(\ref{eq:Mathieu_mod}) and choosing $\delta=0$ for simplicity gives
$$\sum_n\left[-2nc_n-\sum_kc_kc_{n-k}+\delta_{n,0}A-q\delta_{n,1}-q\delta_{n,-1}\right]e^{2inx}\,,$$
where in the following sums run over all integers if not otherwise indicated.
Equating the coefficients of $e^{2inx}$ to zero gives
\begin{eqnarray}
 n&=&0: \sum_kc_kc_{-k}=A\,,\nonumber\\
 n&=&\pm1: c_n=-\frac{1}{2(\pm1+c_0)}\left[q+\sum_{k\neq0,n}c_kc_{n-k}\right]\,,\label{eq:c}\\
 |n|&\geq&2: c_n=-\frac{1}{2(n+c_0)}\sum_{k\neq0,n}c_kc_{n-k}\,.\nonumber
\end{eqnarray}
For $q=0$ this is solved by $c_0=\pm A^{1/2}$, $c_n=0$ for $n\neq0$ and thus one of course obtains the plane phase evolution $\alpha(x)=\pm A^{1/2}x$.

Let us now assume that in the limit $|q|\ll1$ one can neglect $c_n$ for $|n|\geq2$. Then Eq.~(\ref{eq:c}) gives
\begin{eqnarray}
 c_{\pm1}&\simeq&-\frac{q}{2(\pm1+c_0)}\,,\label{eq:c2}\\
 c_0^2&\simeq&A-\frac{q^2}{2(c_0^2-1)}\,.\nonumber
\end{eqnarray}
We are mostly interested in $c_0$ and the second equation con be solved explicitly for $c_0$,
\begin{equation}\label{eq:c_sol}
  c_0^2=\frac{1+A}{2}\pm\frac{1}{2}\left[(1-A)^2-2q^2\right]^{1/2}\,.
\end{equation}
For $|1-A|\gg|q|$ the two solutions can be approximated as
\begin{equation}\label{eq:c_sol2}
  c_0^2\simeq A+\frac{1}{2}\frac{q^2}{|A-1|}\,,\quad c_0^2\simeq 1-\frac{1}{2}\frac{q^2}{|A-1|}\,.
\end{equation}
Thus, the amplitudes are constant and there are only dispersion effects. Only the first solution in Eq.~(\ref{eq:c_sol2})
reproduces the correct solution in the limit $q\to0$, $c_0=\pm A^{1/2}$, so we can discard the second one (it is
probably inconsistent because it leads to divergences $\propto(|A-1|/q)^{|n|}$ for one half of the $c_n$, $n\neq0$ as
one can see from the denominator $\pm1+c_0$.
From Eqs.~(\ref{eq:c}) and~(\ref{eq:c2}) then follows that $c_{1}\simeq\pm q/[2(A-1)]$ and $c_{-1}\simeq\pm q/[2(A+1)]$
which are both smaller than one. Then the higher coefficients become subsequently smaller. Note that if $A$ is not close to
one, $c_{\pm1}$ from Eq.~(\ref{eq:c2}) is of order $|q|$, consistent with what one would expect from Eq.~(\ref{eq:Mathieu2}).
Larger phase shifts of order unity could occur for $|1-A|\simeq|q|$, but for $|q|\ll1$ this will only occur in a very small frequency range.

For $|1-A|\ll|q|$ the two solutions can be approximated as
\begin{equation}\label{eq:c_sol3}
  c_0^2\simeq \frac{1+A}{2}\pm\frac{i}{\sqrt2}|q|\mp\frac{i}{4\sqrt2}\frac{(1-A)^2}{|q|}\,.
\end{equation}
Note that the imaginary parts can give rise to growing modes with an amplitude growth rate $\simeq|q|/(2\sqrt2)$ since
the last term in Eq.~(\ref{eq:c_sol3}) is much smaller than the second term for $|1-A|\ll|q|$. This is known as parametric resonance.
From Eqs.~(\ref{eq:c}) and~(\ref{eq:c2}) then follows $c_1\simeq1/\sqrt2$, $c_{-1}\simeq q/4$ or the other way round
and $c_2\simeq1/4$, $c_{-2}\simeq c_{-1}^2/6\simeq q^2/96$ or the other way round.

\section{Constraints from ALP Over-Densities}
\label{sec:general}
More generally let us now characterise an ALP over-density by its total mass $M$ and radius $R$, assuming spherical symmetry
and a top-hat profile for simplicity. We can then estimate the total mass $\Delta M$ of the ALP star converted during
a time scale $T$. Assuming an isotropic photon flux per unit energy, solid angle and area $j(\epsilon)$, since
the energy width of the resonance is $\Delta\epsilon\simeq m_a|q|/2$, one gets
\begin{equation}\label{eq:Delta_M}
  \Delta M\sim\left[e^{|q|m_aR/\sqrt2}-1\right]\frac{\pi^2}{2}R^2|q|m_a^2j(m_a/2)T\,.
\end{equation}
Similarly to Ref.~\cite{Visinelli:2017ooc} let us now parametrize mass and radius of the ALP over-density by the
dimensionless parameters $\tilde M$ and $\tilde R$,
\begin{eqnarray}
  M&\equiv&\frac{f_a^2}{m_a}\tilde M=\left(\frac{s\alpha_{\rm em}}{2\pi g_{a\gamma}}\right)^2\frac{\tilde M}{m_a}
  \simeq10^{-18}\,\tilde M\left(\frac{f_a}{10^{12}\,{\rm GeV}}\right)^2\left(\frac{\mu{\rm eV}}{m_a}\right)M_\odot\,,\nonumber\\
  R&\equiv&\frac{\tilde R}{m_a}\simeq20\,\tilde R\left(\frac{\mu{\rm eV}}{m_a}\right)\,{\rm cm}\,,\label{eq:M_R}
\end{eqnarray}
where we have used Eq.~(\ref{eq:f_a}). Combining $M\simeq2\pi m_a^2a_0^2R^3/3$ with Eq.~(\ref{eq:Mathieu2a})
for $q$ then yields
\begin{equation}\label{eq:q_abs}
  |q|\simeq\frac{s\alpha_{\rm em}}{2\pi\epsilon_0}\left(\frac{3\tilde M}{2\pi\tilde R^3}\right)^{1/2}\,.
\end{equation}
Inserting into Eq.~(\ref{eq:Delta_M}) gives
\begin{eqnarray}
  \frac{\Delta M}{M}&\sim&\left|\exp\left[\frac{s\alpha_{\rm em}}{2\pi\sqrt2\epsilon_0}\left(\frac{3\tilde M}{2\pi\tilde R}\right)^{1/2}\right]-1\right|
   \frac{\pi^3g_{a\gamma}^2}{\epsilon_0s\alpha_{\rm em}m_a}\left(\frac{3\tilde R}{2\pi\tilde M}\right)^{1/2}m_a^2j(m_a/2)T\label{eq:Delta_M2}\\
 &\sim&3.6\times10^{-28}\exp\left[\frac{s\alpha_{\rm em}}{2\pi\sqrt2\epsilon_0}\left(\frac{3\tilde M}{2\pi\tilde R}\right)^{1/2}\right]
  \left(g_{a\gamma}10^{14}\,{\rm GeV}\right)^2\left(\frac{\mu{\rm eV}}{m_a}\right)\left(\frac{\tilde R}{\tilde M}\right)^{1/2}
  \left(\frac{\left.\epsilon^2j(\epsilon)\right|_{\epsilon=m_a/2}}{10\,{\rm eV}{\rm cm}^{-2}{\rm s}^{-1}{\rm sr}^{-1}}\right)
  \left(\frac{T}{10^{10}\,{\rm y}}\right)\,,\nonumber
\end{eqnarray}
where in the second expression we have assumed that the exponential is much larger than unity since otherwise
there is no significant enhancement, and for the impinging flux we have inserted a typical number applicable to $\mu$eV energies.
We now also see that the parametrization Eq.~(\ref{eq:M_R}) makes the exponent independent of $g_{a\gamma}$ and $m_a$.

For applicability of this simple estimate the profile has to change adiabatically on the scale of the inverse ALP mass,
i.e. $R\gg1/m_a$, or $\tilde R\gg1$ (note that in the box approximation adiabaticity is automatically ensured, except at the boundary).
Furthermore, the growth rate estimate above is valid for $|q|\ll1$ while for a significant enhancement the exponent must be $\gg1$
which also requires $\tilde R\gg1$. Finally, the potential and kinetic energy of the ALPs within the over-density is of order
$G_{\rm N}M/R\simeq(f_a/M_{\rm Pl})^2\tilde M/\tilde R$. This is smaller than the width of the parametric resonance if
$\tilde M\tilde R\lesssim6.4\times10^{-7}(M_{\rm Pl}/f_a)^4$. If this condition is violated and if the over-density does not represent
a condensate in the ground state, the width of the ALP energies may reduce the efficiency of the parametric resonance.

We now note that the energy density of the cosmological diffuse radio background at photon energies $\epsilon\lesssim10^{-6}\,$eV
is about $10^{-10}$ times the dark matter density $\rho_a$, so that any scenario in which more than a fraction $10^{-10}$
of ALPs is converted to radio photons by processes such as the one discussed above, would be ruled out !
More generally, if ALPs constitute a fraction $f_{\rm dm}$ of the dark matter
the fraction $\Delta\rho_a/\rho_a$ of ALPs converted to photons during one Hubble time is constrained by
\begin{equation}\label{eq:constr}
  f_{\rm dm}\frac{\Delta\rho_a}{\rho_a}\lesssim\frac{\Omega_\gamma(m_a/2)}{\Omega_{\rm dm}}\,,
\end{equation}
where $\Omega_\gamma(\epsilon)$ is the energy density of photons per logarithmic energy interval, normalized
to the critical density.

Thus using $\Delta M/M\lesssim10^{-10}$ in Eq.~(\ref{eq:Delta_M2}) for $m_a\lesssim\mu$eV over a Hubble time $t_H\simeq10^{10}\,$y yields
\begin{equation}\label{eq:constr_M_R}
  \tilde M\lesssim5\times10^9\frac{\tilde R}{s^2}\,,
\end{equation}
with additional factors that only depend logarithmically on $g_{a\gamma}$, $m_a$, $\tilde M$, $\tilde R$ and the flux [perhaps include them].
As an example we consider axion mini-clusters which form once the ALP field starts to oscillate at a temperature
$T_{\rm osc}$ given by $H(T_{\rm osc})\simeq m_a$. For a misalignment angle $\theta_{a,0}$ over-densities with radius
$R\sim1/H(T_{\rm osc})\sim1/m_a$ and mass
\begin{equation}
  M_{\rm mini}\sim\rho_a(T_{\rm osc})\frac{4\pi H^{-3}(T_{\rm osc})}{3}\sim\frac{2\pi}{3}\theta^2_{a,0}\frac{f_a^2}{m_a}\,,\label{eq:M_mini}
\end{equation}
where we have used Eq.~(\ref{eq:a0}). This implies $\tilde M\sim2\pi\theta^2_{a,0}/3$, $\tilde R\sim1$ which would
satisfy the constraint Eq.~(\ref{eq:constr_M_R}). For the dilute branch of axion stars Ref.~\cite{Visinelli:2017ooc}
found $\tilde R\simeq(M_{\rm Pl}/f_a)^2/\tilde M$. Inserting into Eq.~(\ref{eq:constr_M_R}) gives
\begin{equation}\label{eq:constr_M_R2}
  \tilde M\lesssim8.6\times10^{11}\frac{1}{s}\left(\frac{10^{12}\,{\rm GeV}}{f_a}\right)\,,
\end{equation}
which is satisfied by the maximum mass of the dilute branch in Ref.~\cite{Visinelli:2017ooc}.
Therefore, axion mini-clusters and axion stars of the type discussed in
Ref.~\cite{Visinelli:2017ooc} seem not to be significantly constrained by these limits.

We can also apply the above constraints to the average Galactic dark matter density. In this case one has $M\simeq4\pi\rho_aR^3/3$,
or
\begin{equation}
  \frac{\tilde M}{\tilde R}\simeq\frac{16\pi^3}{3(s\alpha_{\rm em})^2}\rho_a\left(\frac{g_{a\gamma}}{m_a}\right)^2\tilde R^2
  \simeq1.7\times10^9\left(\frac{g_{a\gamma}10^{14}\,{\rm GeV}}{s}\right)^2
  \left(\frac{\rho_a}{0.3\,{\rm GeV}{\rm cm}^{-3}}\right)\left(\frac{R}{10\,{\rm kpc}}\right)^2\,.\label{eq:M_gal}
\end{equation}
Note that the enhancement factor in Eq.~(\ref{eq:Delta_M2}) then only depends on $g_{a\gamma}$, the ALP density and Galactic scale $R$,
but not on the ALP mass $m_a$ or the parameter $s$. From this we can get a constraint on $g_{a\gamma}$ in the following way: The background
radiation passing through the Galaxy would be enhanced during a time scale $T\sim R\sim10\,$kpc, on the other hand the Galactic
dark matter density is about $10^{15}$ times the energy density in the radio background. Thus we can set $\Delta M/M\lesssim10^{-15}$
with $T\sim10\,$kpc in Eq.~(\ref{eq:Delta_M2}) and use Eq.~(\ref{eq:M_gal}) from which we get
\begin{equation}\label{eq:constr1}
  g_{a\gamma}\lesssim1.8\times10^{-14}\left(\frac{10\,{\rm kpc}}{R}\right){\rm GeV}^{-1}\,.
\end{equation}
This is consistent with our main result Eq.~(\ref{eq:constr1c}).

\section{Effect of Density Profiles on Enhancement, Anisotropy}

If $\rho_a(r)\propto r^{-\alpha}$
close to the galactic center, for $\alpha>2$ the constraint on $g_{a\gamma}$ may become stronger since $|q|r\propto\rho_a^{1/2}r$. To this end
we integrate over an NFW or Burkert dark matter density profile with parameters fitted to Galactic observations \cite{Nesti:2013uwa},
\begin{equation} \label{eq:rho_profile}
    \rho_{_{\rm NFW}}(x) = \frac{\rho_{_{\rm NFW,H}}}{x\left(1+x\right)^2}, \,\qquad  
    \rho_{_{\rm Bur}}(x) = \frac{\rho_{_{\rm Bur,H}}}{\left(1+x\right) \left(1+x\right)^2}.
\end{equation}
Here $x=r/R_{_{\rm H}}$ and $\rho_{_{\rm H}}$ and $R_{_{\rm H}}$ are the scale density and scale radius, respectively, of the fitted model. The scale radius $R_{_{\rm H}}$ in the NFW profile is the radius at which $d \,log \,\rho_{_{\rm NFW}} / d \,log \,r = -2$, whereas in the Burkert profile, it is the radius of the region of constant density.
Using the fitted Galactic values of $\rho_{_{\rm H}}$ and $R_{_{\rm H}}$ from Ref. \cite{Nesti:2013uwa}, NFW: $\rho_{_{\rm H}} = 0.525$ GeV cm$^{-3}$, $R_{_{\rm H}}=16.1$ kpc and Burkert: $\rho_{_{\rm H}} = 1.55$ GeV cm$^{-3}$, $R_{_{\rm H}}=9.26$ kpc, 
we find values of the integral $\int_0^{r_{\rm max}} \rho_a(r)^{1/2} dr$ given in Table~\ref{tab:profile_integrals}.
\begin{table}
 \centering
\begin{tabular}{lcccc}
 \toprule
 \multirow{2}{*}{Density Profile\,\,\,\,\,\,\,\,\,\,\,\,\,} \,\, & \multicolumn{2}{c}{\,\,$r_{\rm max}$= 10 kpc\,\,\,\,\,\,\,\,\,\,\,\,\,\,\,\,} & \multicolumn{2}{c}{$r_{\rm max}=$ 100 kpc} \\
& \,\,I \!\!\!\! & \!\!\! C/AC \,\,& \,\, I \,\, & \,\, C/AC \,\, \\
 \toprule
 NFW     & 2.82 & 22.1 & 5.04 & 3.23\\
 Burkert & 1.89 & 15.1 & 5.86 & 1.79\\
 \toprule
\end{tabular}
\caption{Integral I = $\int_0^{r_{\rm max}} \rho_a(r)^{1/2} dr$ for NFW and Burkert density profiles, in units of the constant product $\left( \rho_a = 0.3 \text{ GeV cm}^{-3}\right)^{-1/2} \times$ 10 kpc. The last column (C/AC) is the ratio of I towards the Galactic center and beyond till $r_{\rm max}$, divided by I towards the anti-center till $r_{\rm max}$, both integrals starting from our position r=8.5 kpc.
}
\label{tab:profile_integrals}
\end{table}
Integrating the resonant enhancement over density profiles implies that the constraint on $g_{a\gamma}$ can  tighten by a factor $\sim 2-3$ or $\sim 5$ when we integrate from the center till 10 kpc or 100 kpc.

An angular anisotropy in the enhanced radio signal is also expected due to our offset from the Galactic center. 
This could be exploited to further tighten constraints on $g_{a\gamma}$ requiring the detected signal to be consistent with the highly isotropic extragalactic radio background \cite{Singal:2017jlh,Holder:2012nm}. The observed upper limits on its fractional anisotropy are $\sim$ 0.01 at arc-minute scales \cite{Holder:2012nm}, ten times smaller than the cosmic infrared background.
On the other hand, the maximum dipolar anisotropy contrast, calculated between the center and anti-center directions using the density profile integrals (Table~\ref{tab:profile_integrals}), is $\sim$ 2-3 for $r_{\rm max}$=100 kpc (and $\sim$ 20 for the case $r_{\rm max}$=10 kpc, close to our Galactocentric radius).

\section{Radio Background Constraints}

We employ observational parameters of existing radio surveys (Table~\ref{tab:radio_obs}) and the infrared and optical background light (Table~\ref{tab:optical_IR}), along with a few proposed surveys with future telescopes, to constrain the axion-photon coupling $g_{a\gamma}$, via Eq.~(\ref{eq:constr1b}). This leads to more specific and detailed constraints over ALP mass windows given by observed wavebands, compared to the in-principle theoretical constraints (red curves in Fig.~\ref{fig:g_m_plot}) continuous over a large range of ALP masses $m_a \lesssim 10$ eV. 
We consider spectral measurements of the extragalactic radio background \cite{Dowell:2018mdb,Singal:2017jlh,Fixsen:2009xn}, which dominates the radio sky (after foreground Galactic synchrotron has been subtracted) for  $\nu \lesssim$ 1 GHz, and for the CMB which is dominant over 1 GHz $\lesssim \nu \lesssim$ 1 THz. 
We also make use of upper limits on the radio sky noise temperature from epoch of reionization (EoR) as well as observed constraints on the large-scale 21 cm power spectrum.  The optical and near infrared background values used are taken from estimates of the lower and upper limits for the background in each band (details mentioned in Table.~\ref{tab:optical_IR}).

We assume that the minimum measurable value of the flux density enhancement factor $f$, in the Rayleigh-Jeans regime, can be taken as 
\begin{equation}
f = (S_{\text{enh}}/ S) \simeq (T+\Delta T)/T = 1 + \Delta T/T
\end{equation}
Here, $S_{\text{enh}}$ is the flux density enhanced by parametric resonance and $S$ the flux density of the background outside of resonance. 
We take $T$ as the brightness temperature of the dominant extragalactic background (excess radio or CMB, assuming foreground removal) at the central frequency $\nu$ of the waveband. 
For $\Delta T$ we employ the quoted r.m.s. noise temperature of that survey or observation, over the channel bandwidth $\Delta \nu$. 
This is equivalent to assuming that the sensitivity to the enhanced signal is set by the r.m.s. noise at that frequency. 
If the signal to noise limit were to be raised by a factor 5, the resultant $g_{a\gamma}$ constraint weakens by at most 2 \% (cf. Eq.~\ref{eq:constr1b}) due to $\Delta T \ll T$ for most radio observations. 
The ratio $\Delta \nu/ \nu \ll 1$ dominates the product ($f\Delta \nu/ \nu$) which appears in a logarithm for the constraint $g_{a\gamma}$. 
Clearly, from Table~\ref{tab:radio_obs}, it is the relative channel bandwidth, at any given frequency, that has significant effect on the $g_{a\gamma}$ constraint. 
Detailed modelling of the impact of foreground contamination and subtraction residuals is beyond the scope of the present work and less important in view of the logarithmic effect on $g_{a\gamma}$.

The radio data and forecasts span the frequency range 10 MHz to 1 THz allowing constraints to be placed on axion condensates over 5 orders of ALP mass 0.08 $\mu \text{ eV} \lesssim m_a \lesssim$ 8000 $\mu \text{ eV}$. The infrared and optical background data span the range  240 $\mu$m $\lesssim \lambda \lesssim$ 0.36 $ \mu$m leading to mass windows in the range $0.01 \text{ eV} \lesssim m_a \lesssim 8 \text{ eV}$.
The mass window in $m_a$ for each observational constraint is determined by the standard waveband of the telescope instrument around each central frequency observed. 
These constraints are shown in Fig.~\ref{fig:g_m_zoom_plot}, where the vertical scale in $g_{a\gamma}$ is magnified cf. Fig.~\ref{fig:g_m_plot}, around the in-principle constraints (red curve) derived from Eq.~(\ref{eq:constr1b}) assuming $f\sim 1$ and $\Delta \nu / \nu \sim 1$. 
Filled regions correspond to constraints from existing observations and dotted lines depict constraints from future forecast observations. 
As in Fig.~\ref{fig:g_m_plot}, we also plot another set of constraints, now 10 times weaker (depicted as darker shaded regions and dashed lines), corresponding to a possibly 10 times shorter extent $R=1$ kpc of the axion condensate.

The $g_{a\gamma}$ constraints range over $1.46 - 2.11 \times 10^{-14}$ GeV$^{-1}$, improving, in some cases, by at most 25 \%, the in-principle constraint. 
The relatively small improvement factor is along entirely expected lines due to the logarithmic dependence on ($f\Delta \nu/ \nu$) in Eq.~(\ref{eq:constr1b}). 
However, it is significant that with radio, infrared and optical observational measurements and limits, we are able to confirm our expected constraint on $g_{a\gamma}$, over several different narrow or broad mass intervals (Fig.~\ref{fig:g_m_zoom_plot}) across the observationally probed ALP mass range $0.08 \mu \text{ eV} \lesssim m_a \lesssim 8 \text{ eV}$. It is fortuitous that cosmic background light estimates are available till optical U-band ($m_a \lesssim 8$ eV) beyond which they become unreliable, while the critical condition (Eq.~\ref{eq:crit_cond_temp}), restricts the condensate ALP mass range to $m_a \lesssim $ 10 eV.

It must be stressed that due to the overall uncertainty on $R$ (the spatial extent of the smooth condensate), the constraints presented will reflect this uncertainty, scaling linearly in $R$, as shown in the two set of curves or regions in Figs.~\ref{fig:g_m_plot} \& \ref{fig:g_m_zoom_plot} for $R$= 10 kpc (lower) or 1 kpc (upper). 
Also, recall that an additional factor of 2-3 improvement in $g_{a\gamma}$ can come from integrating over a realistic dark matter profile $\rho_a(r)$ over 10 kpc.
Nevertheless, we note that these constraints, even at the weaker level of $g_{a\gamma} \lesssim 2 \times 10^{-13}$ GeV$^{-1}$ are approximately 2.5 orders of magnitude stronger than the current helioscope constraints from CAST \cite{CASTAnastassopoulos:2017ftl}, although helioscope constraints don't assume ALPs to constitute dark matter. 
Our condensate constraints for $R$ = 10 kpc are sensitive enough to probe the  $g_{a\gamma}$ values predicted for QCD axion models over the mass range $10^{-5}$ eV $\lesssim m_a \lesssim 10$ eV. 

In applying radio background observational limits to constrain axion-photon coupling, a significant issue we have neglected is the role and mitigation of radio frequency interference (RFI) in actual observations \cite[e.g.,][]{Offringa2012A&A...539A..95O}. A sufficiently bright unresolved spectral line in radio data could be flagged as RFI and excised from the data set so that its detection might be missed. Sensitivity limits and r.m.s. background noise levels are also calculated with RFI removed. However, a spectral feature arising from axion dark matter parametric resonance will be constant and ever-present at that frequency. Time monitoring of RFI variation in the channels being scanned could help to distinguish and characterize possible signals in a spectral search \cite{Offringa2012A&A...539A..95O}. 

\begin{table}
 \centering

  \begin{tabular}
  {lcd{3.6}cccccd{4.2}c}

 \toprule
Telescope/Survey	\,\,&\,\,Reference\,\,\,\,\,\,\,				                        &\text{Frequency}  \!\!\!\!\!\!\!\!\!\!\!\!\!\!\!\!\!\!\!\!	& \,\,Bandwidth	 \,\,& \,\,\,\,\,\,\,$T$\,\,\,\,	 \,\,& \,\,\,\,\,\,$\Delta T$	 \,\,\,\,\,& \,\,\,\,\,\,\,\,\,\,\,\,\multirow{2}{*}{($f\frac{\Delta \nu}{\nu}$)}\,\,\,\,\,\,\,\,\,	& \multirow{2}{*}{$\ln(2f \frac{\Delta \nu}{\nu})$}\!\!\!\!\! \!	& \,\,\,\,\,\text{$m_a$} \!\! \!\!\!\! & \,\,$g_{a\gamma}.10^{14}$ \\
		            &			                                    &\text{(GHz)}\!\!\!\!\!\!\!\!\!\!\!\!\!\!\!	        &(GHz)	    &(K) &(K)       &		            &                  &\,\,\,\,\,\text{($\mu$eV)} \!\!\!\!\!\!\!\!\!  &(GeV$^{-1}$) \\
 \toprule

\multicolumn{10}{c}{\textit{Extragalactic Brightness Temperature Measurements} \cmmnt{(Dowell 2018)} \cite{Singal:2017jlh,Dowell:2018mdb}} \\
\hline
DRAO 10 MHz	            & \cmmnt{Caswell 1976} \cite{Caswell1976MNRAS.177..601C}				                         &0.010 	    &8.00E-06	&85000  &20000	&9.88E-04	&-6.226	&0.08	&1.52 \\
DRAO 22MHz 	            &\cmmnt{Roger 1999} \cite{Roger:1999jy,Costain1969ITAP...17..162C} \cmmnt{Costain 1969 \cite{Costain1969ITAP...17..162C}}			                &0.02225	&3.00E-04	&19212	&4095	&1.64E-02	&-3.420	&0.18	&1.67 \\
LWA LLFSS 40MHz 	            &\cmmnt{Dowell 2017} \cite{Dowell2017MNRAS.469.4537D}				                &0.040052	&9.57E-04	&5792	&963	&2.79E-02	&-2.887	&0.33	&1.71 \\
Japan MU radar	        &\cmmnt{Alvarez 1997} \cite{Alvarez1997AAS..124..315A,Maeda1999AAS..140..145M} \cmmnt{, Maeda 1999 \cite{Maeda1999AAS..140..145M}}                        &0.0465	    &1.65E-03	&4090	&691	&4.15E-02	&-2.489	&0.38	&1.73 \\
LWA LLFSS 50MHz 	            &\cmmnt{Dowell 2017} \cite{Dowell2017MNRAS.469.4537D}				                &0.050005	&9.57E-04	&3443	&526	&2.21E-02	&-3.121	&0.41	&1.71 \\
LWA LLFSS 60MHz 	            &\cmmnt{Dowell 2017} \cite{Dowell2017MNRAS.469.4537D}				                &0.059985	&9.57E-04	&2363	&365	&1.84E-02	&-3.301	&0.50	&1.71 \\
LWA LLFSS 70MHz 	            &\cmmnt{Dowell 2017} \cite{Dowell2017MNRAS.469.4537D}				                &0.070007	&9.57E-04	&1505	&208	&1.56E-02	&-3.470	&0.58	&1.71 \\
LWA LLFSS 80MHz 	            &\cmmnt{Dowell 2017} \cite{Dowell2017MNRAS.469.4537D}				                &0.07996	&9.57E-04	&1188	&112	&1.31E-02	&-3.642	&0.66	&1.71 \\
Haslam 408MHz \cmmnt{Effelsb., Jodr. B. \& Parkes}	&\cmmnt{Haslam 1981} \cite{Haslam1981AA...100..209H,Haslam1982AAS...47....1H,Remazeilles:2014mba}\cmmnt{, 1982 \cite{Haslam1982AAS...47....1H}, Remazeilles 2015 \cite{Remazeilles:2014mba}}	&0.408	&3.50E-03	&15.2	&2.37	&9.92E-03	&-3.920	&3.37	&1.75 \\
Villa Elisa \& Stockert	    &\cmmnt{Reich 1982} \cite{Reich1982AAS...48..219R,Reich1986AAS...63..205R,Reich2001AA...376..861R}\cmmnt{, 1986 \cite{Reich1986AAS...63..205R}, 2001 \cite{Reich2001AA...376..861R}}		                &1.419	&1.55E-02	&3.276	&0.167	&1.15E-02	&-3.774	&11.7	&1.81 \\
ARCADE 2	            &\cmmnt{Fixsen 2011} \cite{Fixsen:2009xn,Singal:2009xq}\cmmnt{, Singal 2011 \cite{Singal:2009xq}}		            &3.15	&2.10E-01	&2.788	&0.045	&6.77E-02	&-1.999	&26.0	    &1.92 \\
ARCADE 2	            &\cmmnt{Fixsen 2011} \cite{Fixsen:2009xn,Singal:2009xq}\cmmnt{, Singal 2011 \cite{Singal:2009xq}}			            &3.41	&2.20E-01	&2.768	&0.045	&6.56E-02	&-2.032	&28.2	    &1.92 \\
ARCADE 2	            &\cmmnt{Fixsen 2011} \cite{Fixsen:2009xn,Singal:2009xq}\cmmnt{, Singal 2011 \cite{Singal:2009xq}}			            &7.97	&3.50E-01	&2.764	&0.06	&4.49E-02	&-2.411	&65.9	    &1.93 \\
ARCADE 2	            &\cmmnt{Fixsen 2011} \cite{Fixsen:2009xn,Singal:2009xq}\cmmnt{, Singal 2011 \cite{Singal:2009xq}}			            &8.33	&3.50E-01	&2.741	&0.062	&4.30E-02	&-2.454	&68.8	    &1.93 \\
ARCADE 2	            &\cmmnt{Fixsen 2011} \cite{Fixsen:2009xn,Singal:2009xq}\cmmnt{, Singal 2011 \cite{Singal:2009xq}}	    	            &9.72	&8.60E-01	&2.731	&0.062	&9.05E-02	&-1.709	&80.3	    &1.97 \\
ARCADE 2	            &\cmmnt{Fixsen 2011} \cite{Fixsen:2009xn,Singal:2009xq}\cmmnt{, Singal 2011 \cite{Singal:2009xq}}			            &10.49	&6.80E-01	&2.731	&0.065	&6.64E-02	&-2.019	&86.7	    &1.96 \\
FIRAS (60-600GHz)	    &\cmmnt{Mather 1994} \cite{Mather:1993ij}				                &60	    &2.10E+01	&2.725	&0.001	&3.50E-01	&-0.356	&496	    &2.11 \\
FIRAS (60-600GHz)	    &\cmmnt{Mather 1994} \cite{Mather:1993ij}				                &250	&2.10E+01	&2.725	&0.001	&8.40E-02	&-1.783	&2070	&2.10 \\
FIRAS (60-600GHz)	    &\cmmnt{Mather 1994} \cite{Mather:1993ij}				                &600	&2.10E+01	&2.725	&0.001	&3.50E-02	&-2.659	&4960	&2.09 \\
\hline
\multicolumn{10}{c}{\textit{Global EoR Experiments}}	\\
\hline
LEDA (30-85MHz)	        &\cmmnt{Price 2018}	\cite{Price2018MNRAS.478.4193P}			                    &0.03	&2.40E-05	&11000	&0.1	&8.00E-04	&-6.438	&0.25	&1.55 \\
LEDA (30-85MHz)	        &\cmmnt{Price 2018}	\cite{Price2018MNRAS.478.4193P}			                    &0.085	&2.40E-05	&800	&0.1	&2.82E-04	&-7.479	&0.70	&1.55 \\
EDGES-2 (50-100MHz)	    &\cmmnt{Bowman 2018} \cite{Bowman:2018yin}			                    &0.050	&6.10E-06	&3300	&0.015	&1.22E-04	&-8.318	&0.41	&1.49 \\
EDGES-2 (50-100MHz)	    &\cmmnt{Bowman 2018} \cite{Bowman:2018yin}				                &0.100	&6.10E-06	&550	&0.015	&6.10E-05	&-9.011	&0.83	&1.49 \\
SARAS-2 (40-200 MHz)	&\cmmnt{Singh 2018} \cite{Singh:2017cnp,Singh:2017gtp}\cmmnt{, 2017 \cite{Singh:2017gtp}}		                &0.110	&1.22E-04	&400	&0.011	&1.11E-03	&-6.111	&0.91	&1.61 \\
SARAS-2 (40-200 MHz)	&\cmmnt{Singh 2018}  \cite{Singh:2017cnp,Singh:2017gtp}\cmmnt{, 2017 \cite{Singh:2017gtp}}					            &0.200	&1.22E-04	&90	    &0.011	&6.10E-04	&-6.709	&1.65	&1.61 \\
\hline
\multicolumn{10}{c}{\textit{21 cm Power Spectrum Constraints}}	\\			\hline
								
LOFAR HBA z=7.9-8.7	    &\cmmnt{Patil 2017}	\cite{Patil:2017zqk}			                    &0.1468	&3.05E-06	&209	&0.1315	&2.08E-05	&-10.088	&1.21	&1.46 \\
LOFAR HBA z=7.9-8.7	    &\cmmnt{Patil 2017}	\cite{Patil:2017zqk}			                    &0.1593	&3.05E-06	&169	&0.1315	&1.92E-05	&-10.169    &1.32	&1.46 \\
LOFAR HBA z=9.6-10.6	&\cmmnt{Patil 2017}	\cite{Patil:2017zqk}			                    &0.1218	&3.05E-06	&338	&0.080	&2.50E-05	&-9.902	    &1.01	&1.46 \\
LOFAR HBA z=9.6-10.6	&\cmmnt{Patil 2017}	\cite{Patil:2017zqk}			                    &0.1343	&3.05E-06	&263	&0.080	&2.27E-05	&-9.999	    &1.11	&1.46 \\
\hline
\multicolumn{10}{c}{\textit{Future Radio Observations}}					\\
\hline
DARE (40-120 MHz)	    &\cmmnt{Burns 2017}	\cite{Burns:2017ndd}			                    &0.040	&5.00E-05	&5800  	&0.06	    &1.25E-03	&-5.991	    &0.33	    &1.58 \\
DARE (40-120 MHz)	    &\cmmnt{Burns 2017}	\cite{Burns:2017ndd}			                    &0.120	&5.00E-05	&300	&0.005	    &4.17E-04	&-7.090	    &0.99	    &1.58 \\
SKA1 (950-1670 MHz)	    &\cmmnt{Yahya 2015}	\cite{Yahya:2014yva}			                    &1.36	&1.00E-05	&3.4	&1.00E-09	&7.35E-06	&-11.127	&11.2	    &1.48 \\
SKA2 (480-1290 MHz)	    &\cmmnt{Yahya 2015}	\cite{Yahya:2014yva}			                    &0.885	&1.00E-05	&4.5	&1.00E-10	&1.13E-05	&-10.698	&7.31	    &1.49 \\
PIXIE (30GHz-6THz)	    &\cmmnt{Kogut 2011}	\cite{Kogut:2011xw}		                            &30	    &1.50E+01	&0.7	&1.00E-07	&5.00E-01	& 0.000	    &248	    &2.10 \\
PIXIE (30GHz-6THz)       &\cmmnt{Kogut 2011}	\cite{Kogut:2011xw}			                        &120	&1.50E+01	&1.34	&1.00E-07	&1.25E-01	&-1.386	    &992	    &2.09 \\
PIXIE (30GHz-6THz)	    &\cmmnt{Kogut 2011}	\cite{Kogut:2011xw}			                        &1000	&1.50E+01	&2.42	&1.00E-07	&1.50E-02	&-3.507	    &8260	&2.06 \\
PRISM (30GHz-6THz)	    &\cmmnt{Andr\'e 2014} \cite{Andre:2013nfa}			                    &30	    &5.00E-01	&0.7	&3.00E-06	&1.67E-02	&-3.401	    &248	    &1.94 \\
PRISM (30GHz-6THz)	    &\cmmnt{Andr\'e 2014} \cite{Andre:2013nfa}	                            &120    &5.00E-01	&1.34	&3.00E-06	&4.17E-03	&-4.787	    &992	    &1.92 \\
PRISM (30GHz-6THz)	    &\cmmnt{Andr\'e 2014} \cite{Andre:2013nfa}				                &1000	&5.00E-01	&2.42	&3.00E-06	&5.00E-04	&-6.908	    &8260	&1.89 \\

\toprule
\end{tabular}
\caption{Radio background parameters for existing and proposed radio survey observations (from the references listed). The implied constraints on $g_{a\gamma}$ as a function of $m_a$ for axion condensate dark matter are calculated using Eq.~(\ref{eq:constr1b}) with $f = 1 + \Delta T/ T$. We have used, from each survey reference, the channel bandwidth value as the spectral resolution $\Delta \nu$ and the r.m.s. noise temperature as $\Delta T$. 
Note that LOFAR parameters are for k = 0.053 hc Mpc$^{-1}$. For PIXIE and PRISM we consider the polarized background taking $T$ equivalent for $\sim$1/10 of the dominant CMB flux and $\Delta T \sim 100$ nK. We assume that PRISM has 30 times smaller bandwidth with 30 times worse sensitivity compared to PIXIE.}
\label{tab:radio_obs}
\end{table}

\begin{table}[]
    \centering
    \begin{tabular}
  {lccccccccd{4.2}c}
\toprule

Instrument/Band	\,\,&\,\,Ref.\,\,\,\,			 &\text{Wavelength}                         &\text{Frequency} 	& \,\,Bandwidth	 \,\,& \,\,$\nu I_{\nu}$	 \,\,& \,\,$\Delta(\nu I_{\nu}$)	 \,\,& \!\!\!\multirow{2}{*}{\,\,($f\frac{\Delta \nu}{\nu}$)}	& \multirow{2}{*}{\,\,\,$\ln(2f \frac{\Delta \nu}{\nu})$}	 & \,\,\,\text{$m_a$} \!\!\!\!\!\!\!\!  & \,\,\,\,\,\,\,\,\,$g_{a\gamma}.10^{14}$ \\
		            &  \!\!\!\!\!\!\!                                   &\text{($\mu$m)} &\text{(GHz)}        &(GHz)	    &\multicolumn{2}{c}{\!\!\!\!(nW\,m$^{-2}$\,sr$^{-1}$)}       &	 &                  &\,\text{(eV)} \!\!\!\!\!\!\!\!\!      &\,\,\,\,\,\,(GeV$^{-1}$) \\
\toprule
HDF U-band		&\cite{Madau2000MNRAS.312L...9M,HESS2013AA...550A...4H,Meyer2012AA...542A..59M}  &0.36	&833	&151.7	&2.87	&4	&0.436	&-0.138	&6.89	&1.95\\
HDF B-band		&\cite{Madau2000MNRAS.312L...9M,HESS2013AA...550A...4H,Meyer2012AA...542A..59M}  &0.45	&667	&133.1	&4.57	&5	&0.418	&-0.179	&5.51	&1.94\\
HDF R-band		&\cite{Madau2000MNRAS.312L...9M,HESS2013AA...550A...4H,Meyer2012AA...542A..59M}  &0.67	&448	&107.1	&6.74	&10	&0.594	&0.172	&3.70	&1.93\\
HDF I-band		&\cite{Madau2000MNRAS.312L...9M,HESS2013AA...550A...4H,Meyer2012AA...542A..59M}  &0.81	&370	&71.06	&8.04	&15	&0.550	&0.095	&3.06	&1.92\\
HDF J-band		&\cite{Madau2000MNRAS.312L...9M,HESS2013AA...550A...4H,Meyer2012AA...542A..59M}  &1.1	&273	&50.00	&9.71	&17	&0.504	&0.009	&2.26	&1.91\\
HDF H-band		&\cite{Madau2000MNRAS.312L...9M,HESS2013AA...550A...4H,Meyer2012AA...542A..59M}  &1.6	&188	&35.47	&9.02	&15	&0.504	&0.007	&1.55	&1.89\\
HDF K-band		&\cite{Madau2000MNRAS.312L...9M,HESS2013AA...550A...4H,Meyer2012AA...542A..59M}  &2.2	&136	&21.83	&7.92	&14	&0.443	&-0.121	&1.13	&1.87\\
\hline
IRAC 3.6 $\mu$m	&\cite{Fazio2004ApJS..154...39F,Savage2005astro.ph.11359S}  &3.6	&83.3	&17.90	&5.4	&12	&0.692	&0.325	&0.69	&1.87\\
IRAC 4.5 $\mu$m	&\cite{Fazio2004ApJS..154...39F,Savage2005astro.ph.11359S}  &4.5	&66.7	&15.24	&3.5	&6	&0.620	&0.216	&0.55	&1.86\\
IRAC 5.8 $\mu$m	&\cite{Fazio2004ApJS..154...39F,Savage2005astro.ph.11359S}  &5.8	&51.7	&13.44	&3.6	&5	&0.621	&0.216	&0.43	&1.85\\
IRAC 8.0 $\mu$m	&\cite{Fazio2004ApJS..154...39F,Savage2005astro.ph.11359S}  &8.0	&37.5	&14.39	&2.6	&4	&0.974	&0.667	&0.31	&1.85\\
\hline
MIPS 24 $\mu$m	&\cite{Bethermin2010AA...512A..78B}\cmmnt{Béthermin et al. (2010)}	&23.7	&12.7	&2.54	&2.86	&4	&0.480	&-0.040	&0.105	&1.78\\
MIPS 70 $\mu$m	&\cite{Bethermin2010AA...512A..78B}\cmmnt{Béthermin et al. (2010)}	&71	    &4.23	&1.15	&6.6	&10	&0.685	&0.315	&0.035	&1.75\\
\hline
DIRBE 100 $\mu$m	&\cite{Berta2011AA...532A..49B,Dole2006AA...451..417D,Hauser1998ApJ...508...25H}\cmmnt{Dole 2006}	&100	&3.00	&0.97	&14.4	&14	    &0.640	&0.247	&0.025	&1.73\\
DIRBE 140 $\mu$m	&\cite{Berta2011AA...532A..49B,Dole2006AA...451..417D,Hauser1998ApJ...508...25H}\cmmnt{Dole 2006}	&140	&2.14	&0.61	&12	    &6.9	&0.445	&-0.117	&0.018	&1.71\\
DIRBE 240 $\mu$m  	&\cite{Berta2011AA...532A..49B,Dole2006AA...451..417D,Hauser1998ApJ...508...25H}\cmmnt{Dole 2006}	&240	&1.25	&0.5	&12.3	&2.5	&0.476	&-0.048	&0.010	&1.69\\
\toprule
    \end{tabular}
    \caption{Parameters describing measurements of the optical and infrared extragalactic background light (from the references listed). Constraints on $g_{a\gamma}$ as a function of $m_a$ for axion condensate dark matter are calculated using Eq.~(\ref{eq:constr1b}) with $f = S_{\text{enh}}/S \approx 1 + \Delta S/S$. We have used the band width as the spectral resolution $\Delta \nu$. The background flux $S$ or intensity $\nu I_{\nu}$ is adopted from lower limits (integrated counts) or detections, whereas the maximum enhancement over background $\Delta S$ or $\Delta(\nu I_{\nu})$ is taken from upper limits  from gamma-ray opacity \cite{HESS2013AA...550A...4H,Meyer2012AA...542A..59M} or fluctuation analysis \cite{Savage2005astro.ph.11359S}. Where upper limits were not found (140 and 240 $\mu$m), measurement uncertainties were used.}
    \label{tab:optical_IR}
\end{table}

\end{widetext}

\end{document}